\documentclass[final,useamsfonts]{pasj00}

\usepackage{graphicx}
\usepackage{color}
\usepackage{ulem}

\newcommand{\etacar}{$\eta$~Car}

\input{macro.inp}

\SetRunningHead{Hamaguchi et al.}{\SUZAKU\ Observation of \etacar}
\Received{2006 August 10}
\Accepted{2006 September 29}
\Published{}

\begin{document}

\title{\SUZAKU\  Observation of Diffuse X-ray Emission from\\ the Carina Nebula}

\author{Kenji \textsc{Hamaguchi}\altaffilmark{1,2}, Robert \textsc{Petre}\altaffilmark{1},
Hironori \textsc{Matsumoto}\altaffilmark{3}, Masahiro \textsc{Tsujimoto}\altaffilmark{4},\\
Stephan S. \textsc{Holt}\altaffilmark{5}, Yuichiro \textsc{Ezoe}\altaffilmark{6},
Hideki \textsc{Ozawa}\altaffilmark{7}, Yohko \textsc{Tsuboi}\altaffilmark{8},\\
Yang \textsc{Soong}\altaffilmark{1,2},
Shunji \textsc{Kitamoto}\altaffilmark{4}, Akiko \textsc{Sekiguchi}\altaffilmark{4}, 
Motohide \textsc{Kokubun}\altaffilmark{9}
}

\altaffiltext{1}{Astrophysics Science Division, NASA Goddard Space Flight Center,
Greenbelt, MD \\
20771, USA}
\altaffiltext{2}{Universities Space Research Association, 
10211 Wincopin Circle, Suite 500, Columbia, MD\\
 21044-3432, USA}
\altaffiltext{3}{Department of Physics, GraduateSchool of Science, Kyoto University, Kita-Shirakawa,\\
Sakyo-ku, Kyoto 606-8502}
\altaffiltext{4}{Department of Physics, Faculty of Science, Rikkyo University, 3-34-1  Nishi-Ikebukuro,\\
Toshima-ku, Tokyo 171-8501}
\altaffiltext{5}{Franklin W. Olin College of Engineering, Needham, MA 02492, USA}
\altaffiltext{6}{The Institute of Space and Astronautical Science, 
3-1-1 Yoshinodai, Sagamihara, \\
Kanagawa 229-8510}
\altaffiltext{7}{Graduate School of Science Earth and Space Science
Osaka University,\\ 
Machikaneyama 1-1, Toyonaka Osaka, 560-0043}
\altaffiltext{8}{Department of Physics, Faculty of Science and Engineering, Chuo University, 1-13-27\\ Kasuga, Bunkyo-ku, Tokyo 112-8551}
\altaffiltext{9}{Department of Physics, University of Tokyo, 7-3-1 Hongo, Bunkyo-ku, Tokyo 113-0011}
\email{kenji@milkyway.gsfc.nasa.gov}
\maketitle

\begin{abstract}
We studied extended X-ray emission from the Carina Nebula taken with 
the \SUZAKU\  CCD camera XIS on 2005 Aug. 29.
The X-ray morphology, plasma temperature and absorption to the plasma are
consistent with the earlier \EINSTEIN\ results.
The \SUZAKU\ spectra newly revealed emission lines from various species including oxygen,
but not from nitrogen.
This result restricts the N/O ratio significantly low, compared with evolved massive stellar winds,
suggesting that the diffuse emission is originated in an old supernova remnant or a 
super shell produced by multiple supernova remnants.
The X-ray spectra from the north and south of \etacar\  showed
distinct differences between 0.3$-$2~keV.
The south spectrum shows strong L-shell lines of iron ions and K-shell lines of silicon ions,
while the north spectrum shows them weak in intensity.
This means that silicon and iron abundances are a factor of 2$-$4 higher in the 
south region than in the north region.
The abundance variation may be produced by an SNR ejecta, or relate to the
dust formation around the star forming core.
\end{abstract}

\section{Introduction}
\label{sec:intro}

Not a few giant H\emissiontype{II} regions emit extended soft X-rays
with \KT\  $\sim$0.1--0.8~keV, log \LX\  $\sim$33--35~\UNITLUMI, and extent $\sim$1--10$^{3}$~pc
(RCW~38: \cite{Wolk2002}, M17: \cite{Townsley2003}, NGC~6334: \cite{Ezoe2006}, Carina Nebula: \cite{Seward1979,Seward1982,Evans2003},
extragalactic H\emissiontype{II} regions: \cite{Kunz2003}, \cite{Strickland2004}).
This high surface brightness cannot be accounted for by an extrapolation of the discrete
source luminosity function, and thus is thought to arise from diffuse plasma.
The required plasma temperature and thermal energy can be produced by 
collisions or termination of fast winds
from main-sequence or embedded young O stars \citep{Seward1982,Townsley2003,Ezoe2006},
but the extended emission is often observed from regions apart from massive stellar clusters.
Another origin, such as an unrecognized supernova remnant (SNR), cannot be ruled out.

In principle, the origin of the diffuse emission can be determined by measuring its composition.
For example, the plasma should be overabundant in nitrogen and neon if it originates from
winds from a nitrogen-rich Wolf-Rayet star (WN),
while it would be overabundant in oxygen if it arises from a Type~II SNR.
The temperature of the plasma, typically a few million degrees, make soft X-ray band studies
highly desirable, because of the presence in this band of strong lines from these elements, plus
carbon, silicon and iron.

The Carina Nebula, which contains several evolved and main-sequence massive stars
such as \etacar, WR~25 and massive stellar clusters such as Trumpler~14 (Tr 14),
emits soft diffuse X-rays 10--100 times stronger
than any other Galactic giant H\emissiontype{II} region 
(\LX\  $\sim$10$^{35}$~\UNITLUMI, \cite{Seward1979}).
The high surface brightness made possible the discovery of the diffuse emission by
the \EINSTEIN\  Observatory in the late 1970's.
The \EINSTEIN\  observations revealed that
the diffuse emission tends to be associated with optically bright regions containing massive stars, 
and is strongest in regions bordering the western part of the V-shaped dust lane
\citep{Seward1982}. 
Recent \CHANDRA\  observations provided a point source free measurement of the diffuse flux  \citep{Evans2003}, and
suggested 
presence of a north-south Fe and Ne abundance gradient \citep{Townsley2006}.

No detailed abundance measurement of the
diffuse emission from the Carina Nebula has been made.
The X-ray CCD cameras (XISs: X-ray Imaging Spectrometer) onboard the \SUZAKU\  observatory
have the best spectral resolution for extended soft X-ray emission.
It is thus suitable for measuring the abundance of key elements in the diffuse plasma.
To compensate for the limited angular resolution of \SUZAKU, 
we have also utilized the archival \XMM\  \citep{Jansen2001} and \CHANDRA\  \citep{Weisskopf2002} 
data from the Carina Nebula to
investigate the spatial distribution of the diffuse emission and the contamination from X-ray point sources.

\section{Observations \& Data Reduction}
\label{sec:obs}

\SUZAKU\  (a.k.a. Astro-E2, \cite{Mitsuda2006}) is the 5th Japanese X-ray observatory, developed by a
Japanese-US collaboration, and launched on 2005 July 10.
The observatory is equipped with five thin-foil
X-Ray Telescopes (XRT) with half power diameter (HPD) of $\sim$2$'$
\citep{Serlemitsos2006}.
At their focal planes are
four CCD focal plane detectors (XIS, \cite{Koyama2006}) and an X-ray calorimeter
(XRS, \cite{Kelly2006}),
which unfortunately malfunctioned a month after the launch.
The observatory also has a hard X-ray detector (HXD, \cite{Takahashi2006,Kokubun2006}).
The XIS comprises four CCD cameras, XIS0 through XIS3, three of which (XIS 0, 2 and 3)
use front-illuminated (FI) CCD chips while XIS1 utilizes the
back-illuminated (BI) technology.
Compared with CCD detectors on earlier X-ray observatories,
both the BI and FI chips have better spectral resolution at energies below $\sim$1~keV:
their FWHM energy resolution at 0.5~keV is $\sim$40~eV (FI) and $\sim$50~eV (BI),
with negligible low-energy tails.

\begin{figure*}
\begin{center}
\FigureFile(160mm,79mm){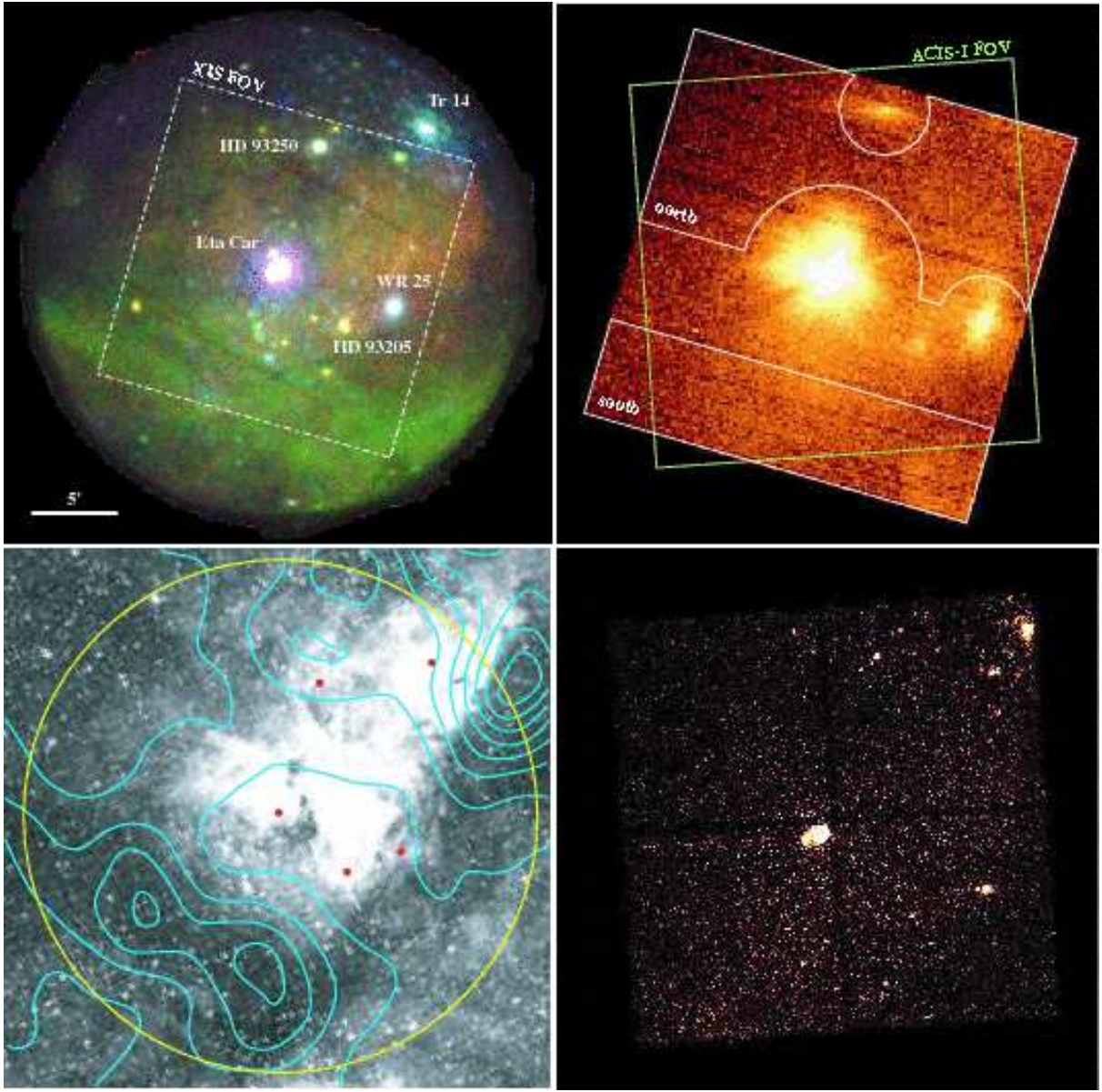}
\end{center}
\caption{{\it Top left}: \XMM\  EPIC MOS image color-coded to represent the soft band
(0.2--0.6~keV) to red, medium band (0.6--1.2~keV) to green and hard band (1.2--10~keV)
to blue. 
The bar-dot lines show the XIS \FOV\  in the 2005 Aug. 29 observation.
{\it Top right}: \SUZAKU\  BI image in the 2005 Aug. 29 observation.
The solid white lines show event extraction regions, and
the solid green lines show the \CHANDRA\  \FOV\  in the 1999 Sep. 6 observation.
{\it Bottom left}: Integrated intensity map of the $^{12}$CO emission (sky blue contour) 
on the optical image taken from the Digitized Sky Survey (gray scale image),
which is a part of Figure 1 in \citet{Yonekura2005}.
The red spots show the bright X-ray sources introduced in the {\it top left} image.
The yellow circle shows the \XMM\  EPIC MOS \FOV.
{\it Bottom right}: \CHANDRA\  image in the 1999 Sep. 6 observation.
All the X-ray images are drawn with logarithmic scale and
do not correct the vignetting effect.
\label{fig:xmmsuzakuimage}}
\end{figure*}

\SUZAKU\  observed the Carina Nebula twice, on 2005 Aug. 29 and 2006 Feb. 3, 
with similar pointings, putting \etacar\  at the center of the XIS \FOV\  
(Figure~\ref{fig:xmmsuzakuimage}).
In both observations, the XISs were operated with the Normal mode.
The first observation was 16 days after the XIS first light (2005 Aug 13)
when contamination on the optical blocking filter was almost negligible,
while the second observation, 174 days after the first light, 
suffered significant degradation of the soft response
by progressive contamination on the XIS
of a factor of $\sim$2 between 0.4--0.5~keV on axis.
We thus did not analyze the second observation in this paper.
We also did not use the HXD data, whose X-ray signals are mostly from \etacar.

We used the 
HEAsoft\footnote{http://heasarc.gsfc.nasa.gov/docs/software/lheasoft/} analysis package ver.~6.0.6
for data analysis.
For the version~0.7\footnote{
The version 0.x process uses the earliest detector calibration and \SUZAKU\  software still under development,
and was released only to the Suzaku team members (Science Working Group: SWG)
to verify the data quality.
The process does not perform some detailed calibration of the data, 
such as correction of the satellite wobbling
and event time tagging optimized for the XIS operational mode.
The latest processing as of 2006 July is version 0.7,
which achieved an absolute energy scale within $\pm5$~eV below 1 keV
and $\pm$0.2\% at the iron K$\alpha$ energy for the XIS data.
} unscreened data,
we screened out hot and flickering pixels with {\tt cleansis} 
implemented in the {\tt xselect} and removed exposure frames with telemetry saturation.
We excluded the data taken during the South Atlantic Anomaly (SAA) passage 
and within 256~sec after the passage.
We selected data taken while the geomagnetic cut-off rigidity
(COR) was larger than 4 GV,
elevation from above the night-time Earth horizon (ELV) was larger than 5\DEGREE, and 
elevation above the sun-lit Earth rim (DaY Earth ELeVation: DYE\_ELV) was larger than 10\DEGREE\  for the FI data and 20\DEGREE\  for the BI.
The tight DYE\_ELV filtering criterion for the BI was intended to minimize
contamination by neutral nitrogen and oxygen lines from the sunlit terrestrial atmosphere.
There were no remarkable particle background flares that degraded the data quality.
The net exposure of the each FI chip and the BI were $\sim$60~ksec and 49.0~ksec, respectively.
In spectral fits, we used the XIS response matrices released on 2006-02-13
and generated auxiliary files with {\tt xissimarfgen-2006-05-28}, 
which considers additional absorption due to the XIS contamination.

\begin{figure*}[t]
\begin{center}
\FigureFile(160mm,131.1mm){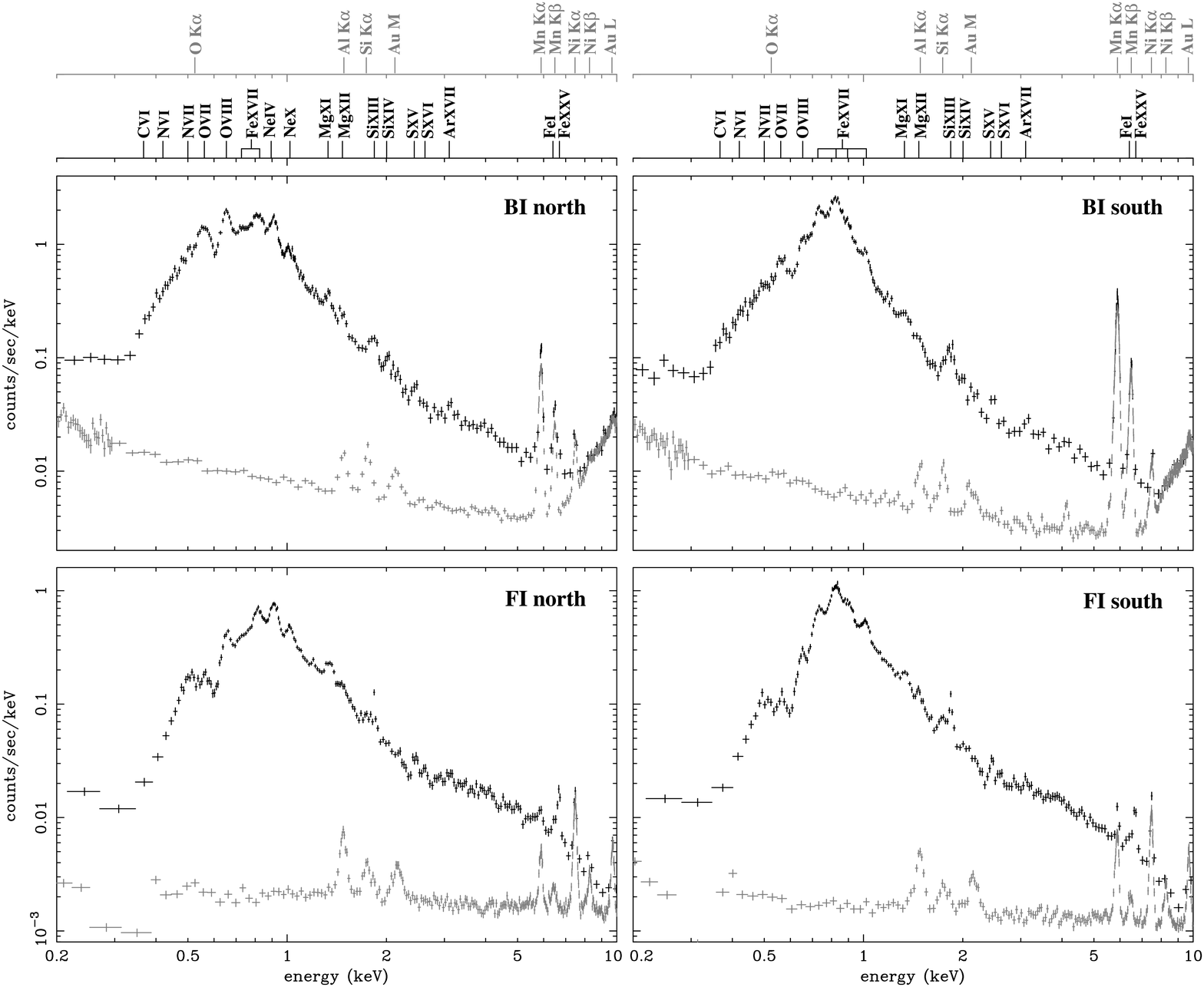}
\end{center}
\caption{
BI ({\it top}) and FI ({\it bottom}) spectra of the north ({\it left}) and south ({\it right}) regions.
The black shows the source spectra and the grey shows the detector internal background,
obtained from night Earth observations.
The above labels demonstrate energies of emission lines detected or concerned
with this result.
\label{fig:suzakudiffspec}
}
\end{figure*}

\section{Spatial Distribution of the X-ray Sources}
\label{sec:spatialdist}

To understand the spatial distribution of the diffuse X-ray emission and X-ray point sources in the field,
we surveyed archival \XMM\  data-sets aimed at \etacar.
There are in total 11 observations performed between 2000 and 2003\footnote{
Sequence ID: 112580601, 112580701, 145740101,
145740201, 145740301, 145740401, 145740501, 160160101, 160160901, 145780101, 
160560101, 160560201, 160560301.}.
Figure~\ref{fig:xmmsuzakuimage} shows a true color image
accumulated from all the MOS data taken with the prime full window mode
({\it red}: 0.2$-$0.6~keV, {\it green}: 0.6$-$1.2~keV, {\it blue}: 1.2$-$10~keV).
The image depicts several bright X-ray point sources:
\etacar\  (LBV star), WR25 (Wolf-Rayet star), HD~93250 and HD~93205 (O3 stars),
and Tr~14 (massive stellar cluster),
as well as multiple weak point sources especially concentrated between \etacar\  and WR~25.
The image also clearly shows two kinds of apparently extended emission, seen as red and green.
The ``red" emission --- relatively strong below 0.6~keV --- is located between
Tr~14, WR~25 and \etacar.
As is suggested by \citet{Seward1982},
the emission is stronger at the west side with a sharp cut-off at the boundary 
of the V-shaped optical dark lane.
The ``green" emission --- relatively strong above 0.6~keV --- is located to the south of \etacar,
extending east-west, where another optical dark lane and CO gas condensation are located
(\cite{Yonekura2005}, also see Figure~\ref{fig:xmmsuzakuimage}).

The \SUZAKU\  XIS \FOV\  (Figure~\ref{fig:xmmsuzakuimage})
covers most of the X-ray point and extended sources detected in the \XMM\  image.
Among the point sources,
the XIS image resolves four bright X-ray sources, \etacar, WR~25, HD~93250 and HD~93205.
To minimize contamination,
we excluded regions around these sources,
within 4\ARCMIN\  from \etacar\  and 2\ARCMIN\  from the other three sources.
We then defined the ``north" and ``south" regions as in Figure~\ref{fig:xmmsuzakuimage},
to extract spectra of the ``red" and ``green" emission, respectively.
We excluded an area between the north and south regions containing multiple weak point sources.
Bright spots on the BI image at the upper and lower right corners are $^{55}$Fe calibration sources.
We included these regions in the BI data to increase the photon statistics for the diffuse emission,
since the calibration sources do not significantly contaminate the spectrum outside 5.9 and 6.4~keV,
while we removed them from the FI data so we could investigate the Fe K line profile around 6$-$7~keV.

\section{Characteristics of the North and South Spectra}

We extracted all the XIS0-3 spectra from both the north and south regions and
combined the spectra of the FI chips (XIS0, XIS2 and XIS3), whose spectral responses 
are almost identical. (Hereafter, we call the combined spectrum as "FI spectrum".)
We overlaid these FI and BI spectra of the north and south regions 
on spectra accumulated from observations of the night side of the Earth 
(Figure~\ref{fig:suzakudiffspec}).
Events detected during the night Earth observations originate from particle background,
X-ray fluorescence of Ni, Au and Si inside the camera body, $^{55}$Fe calibration source
and detector electronic noise.
Accurate reproduction of a background spectrum requires proper accounting for
the range of COR values encountered during the observation.  For each of the two
regions, we accumulated in detector coordinates for each field 
night Earth spectra corresponding to various COR values
and constructed a spectrum from a weighted average of these.
The night Earth spectra match with the observed spectra at 
 $\gtrsim$10~keV for the FI with less than $\sim$7.5\% discrepancy, and $\gtrsim$8~keV for the BI
with less than $\sim$2.5\% discrepancy.
A small discrepancy by $\sim$10\% is seen in lines from the $^{55}$Fe calibration source,
which decayed from the Carina Nebula observation on 2006 Aug 29 to the span
between 2005 September -- 2006 May when the night Earth spectra were accumulated.
This means that the internal background during this observation is well reproduced with the night Earth spectra, and we therefore use the night Earth data as a background in subsequent analysis.
Both the north and south spectra show substantial excess X-ray emission above the background level
between 0.2$-$10~keV, including prominent emission lines.

The north and south spectra show similar spectral features above 2~keV and below 0.3~keV.
Above 2~keV, both spectra show emission lines from 
S\emissiontype{XV}, S\emissiontype{XVI}, Ar\emissiontype{XVII} and 
Fe\emissiontype{XXV} ions, emitted from hot plasma
with \KT\  $\gtrsim$2~keV,
as well as a marginal detection of fluorescence from cold iron at 6.4~keV.
The south region has $\sim$60$-$70\% of the number of events from the north region,
which just corresponds to the ratio of the effective area, and does 
not strongly depend on the photon energy.
This means that the flux level and continuum slope are also similar between the north and south spectra
above 2~keV.
This similarity is seen for emission below 0.3~keV as well, though it is only measurable with the BI.
The north and south spectra show no remarkable variation below 0.3~keV and above 2~keV.

\begin{figure}
\FigureFile(80mm,59.75mm){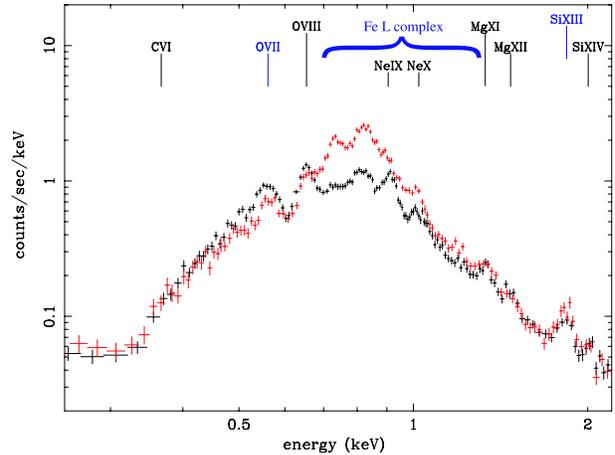}
\caption{North (black) and south (red) spectra between 0.3$-$2 keV overlaid.
The north spectrum is normalized by 64\%, to adjust the effective area of the south region at 2~keV.
The above labels show detected emission lines,
among which blue elements exhibit distinct differences between the north and south spectra.
\label{fig:carnebnorthsouthoverlay}
}
\end{figure}

In contrast, the spectra between 0.3$-$2 keV are conspicuously different.
Figure~\ref{fig:carnebnorthsouthoverlay} shows an overlay of the BI spectra between 0.3--2~keV,
with the north spectrum normalized by 64\% to compensate for the effective area difference
at 2~keV. 
A strong difference is seen between 0.7~keV and 1.2~keV, which
apparently is the source of the two colors of diffuse emission in Figure~\ref{fig:xmmsuzakuimage}.
The band in which the difference is found is dominated by emission lines from the iron L-shell complex.
Additionally, the south spectrum shows a stronger Si\emissiontype{XIII} line and possibly a stronger C\emissiontype{VI} line at $\sim$0.39~keV, while 
the north spectrum shows a clear excess at and below the O\emissiontype{VII} line.
The south spectrum shows no clear evidence of lines from Ne\emissiontype{IX} or 
Ne\emissiontype{X} ions, but these could be hidden by the Fe L-shell complex.
Except for the energy bands with these emission lines, 
the spectral shapes look similar.
This suggests that the differences represent an
elemental abundance variation, 
and not a temperature difference.
We should note that 
though a small marginal peak at 0.5~keV looks like an emission line of hydrogen-like nitrogen,
it is more likely produced by a combination of interstellar absorption
and the instrumental oxygen edge at 0.54~keV.

These results suggest that the spectra from both regions are roughly divided into three components:
emission above 2~keV (hard component, hereafter HC) and below 0.3~keV (soft component, SC),
which do not differ significantly between the north and south regions; and
emission between 0.3 and 2~keV (medium component, MC) with
remarkable variation in some emission lines.
We see strong variation in emission lines from the Fe L-shell complex in the MC,
but not even slightly in emission lines from Fe\emissiontype{XXV} ions in the HC.
This supports the idea that each component has a different origin.
The most plausible origin of the SC is foreground emission; for the HC it is 
a combination of point sources, Cosmic X-ray Background (CXB) radiation, 
and Galactic Ridge X-ray Emission (GRXE); and for the MC, truly diffuse emission from the
Carina Nebula.
In the next section, we estimate these contributions in detail.

\section{Contamination from X-ray Sources}
\label{sec:contami}

\begin{figure*}
\begin{center}
\FigureFile(80mm,58.97mm){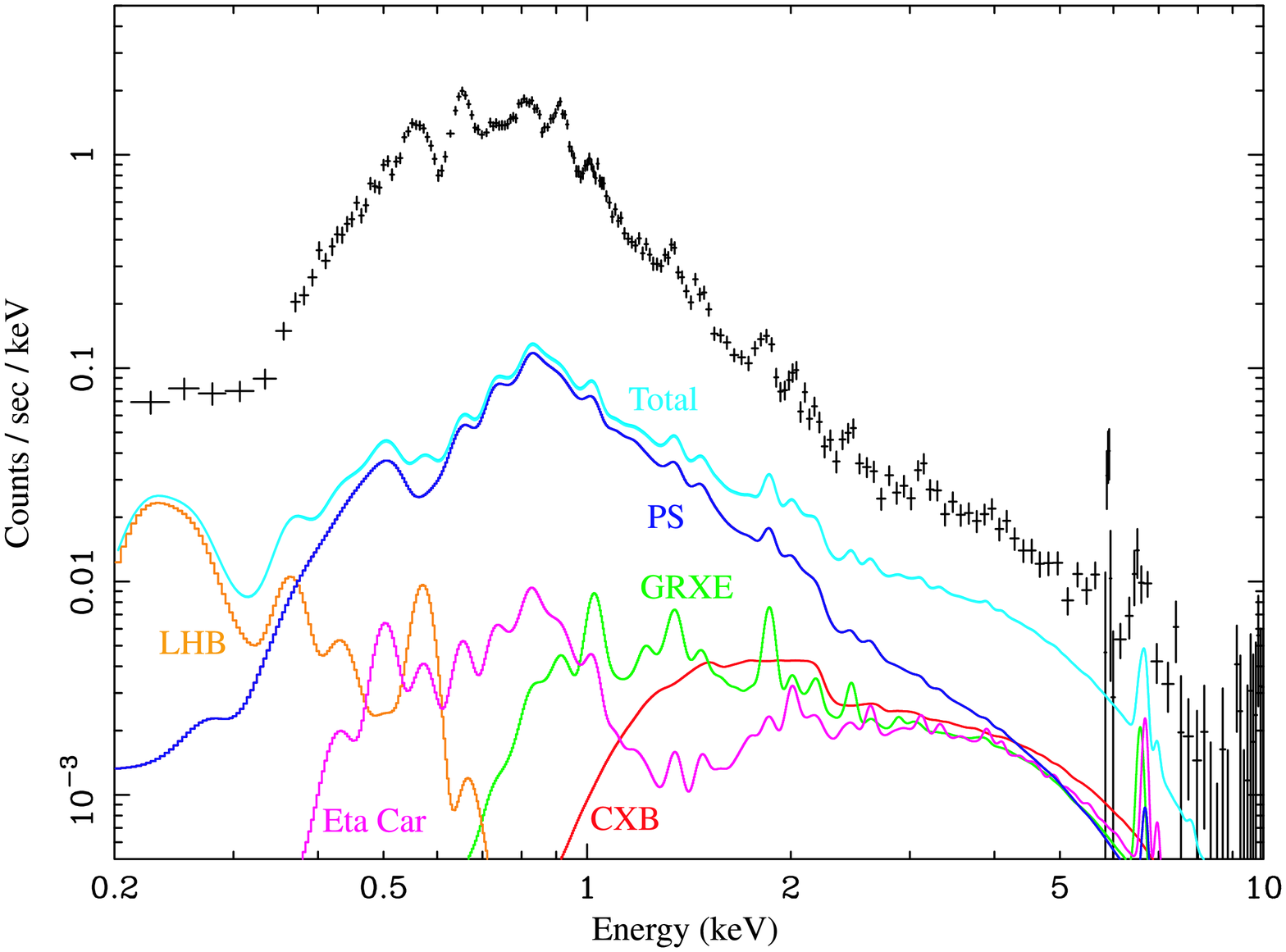}
\FigureFile(80mm,58.97mm){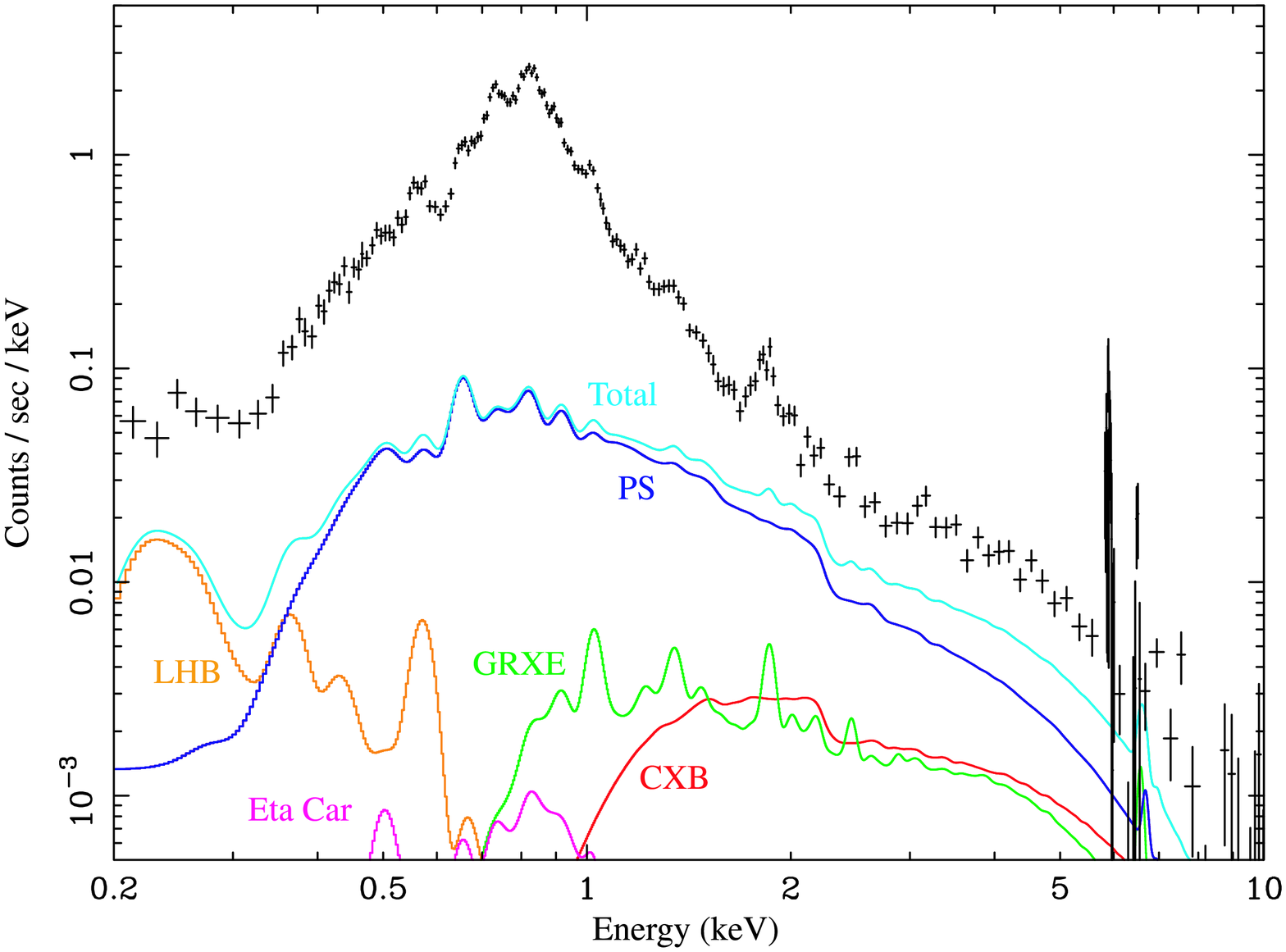}
\end{center}
\caption{Background subtracted BI spectra ({\it left}: north, {\it right}: south)
with possible spectra of contamination sources.
The point source spectra are scaled to the area coverage of the
\SUZAKU\ source regions (17\% to the north and 59\% to the south).
\label{fig:suzakuspecwcontami}
}
\end{figure*}

Both spectra include emission from numerous point sources that are not resolved 
with \SUZAKU\  (see the \XMM\  image in Figure~\ref{fig:xmmsuzakuimage}),
CXB, GRXE and Local Hot Bubble (LHB) emission.
In this section, we estimate the contribution of these contaminants in the \SUZAKU\  spectra.
We plot their estimated contribution in Figure~\ref{fig:suzakuspecwcontami},
using point or extended source responses generated with {\tt xissimarfgen-2006-05-28}.

\subsection{Point Sources}
\label{subsec:psestimate}

In a \CHANDRA\  observation in 1999,
\citet{Evans2003} detected 25 and 12 X-ray point sources from the north and south regions, respectively.
They estimated the X-ray flux of each detected source from its count rate,
assuming a typical stellar spectrum and extinction to the Carina Nebula
(log $T$ = 6.65 $K$, \NH\  = 3$\times$10$^{21}$~\UNITNH).
However, 
several point sources in the field are found to have significantly higher plasma temperature
than the typical stellar coronal source \citep{Colombo2003,Evans2004}.
This means that representing the contribution of unresolved point sources in the \SUZAKU\  
spectra using a typical stellar spectrum scaled to the total point source flux calculated
from tables in \citet{Evans2003} would be incorrect.
We therefore analyzed the archival \CHANDRA\  observation of the Carina Nebula.

\begin{figure*}
\begin{center}
\FigureFile(80mm,53.04mm){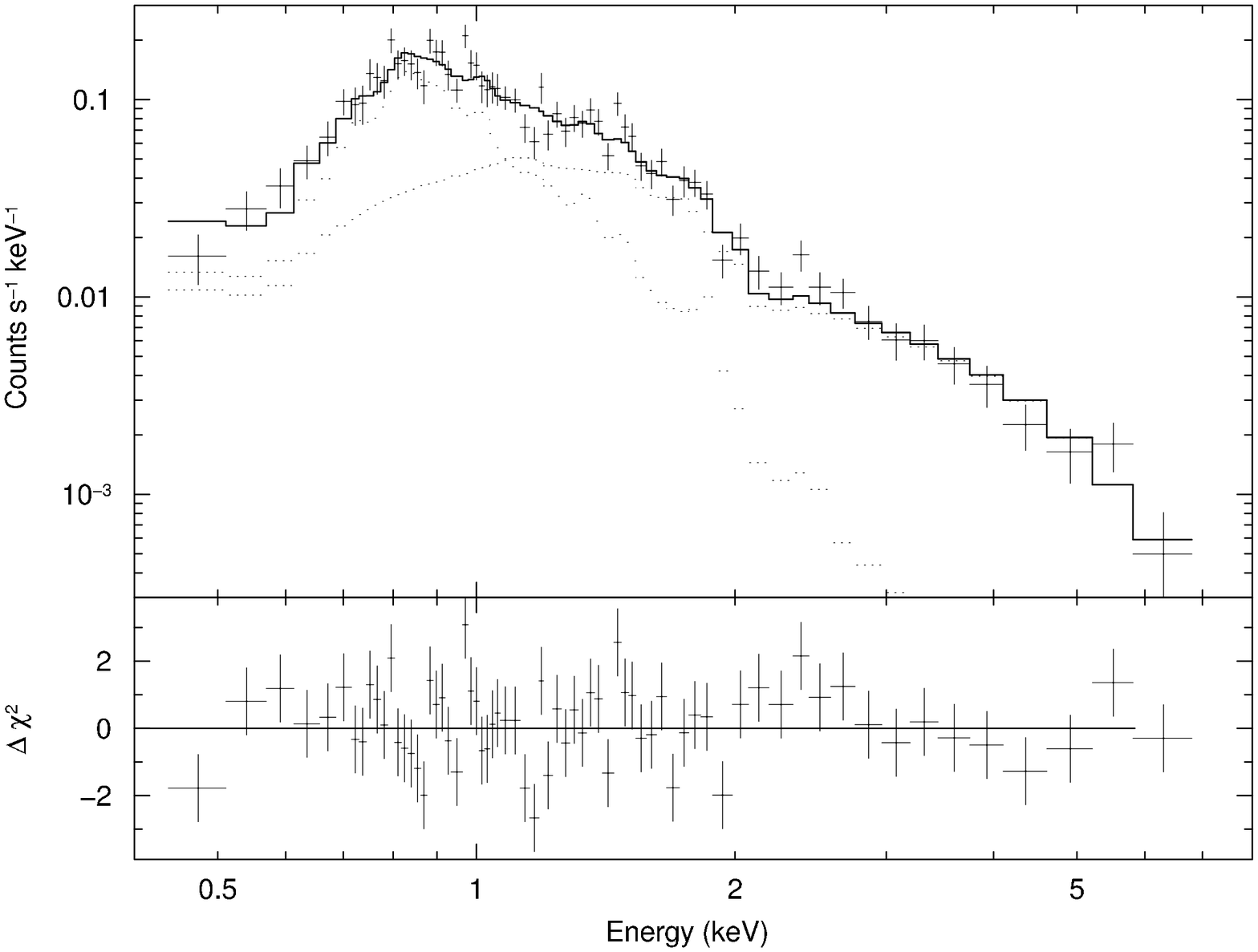}
\FigureFile(80mm,53.04mm){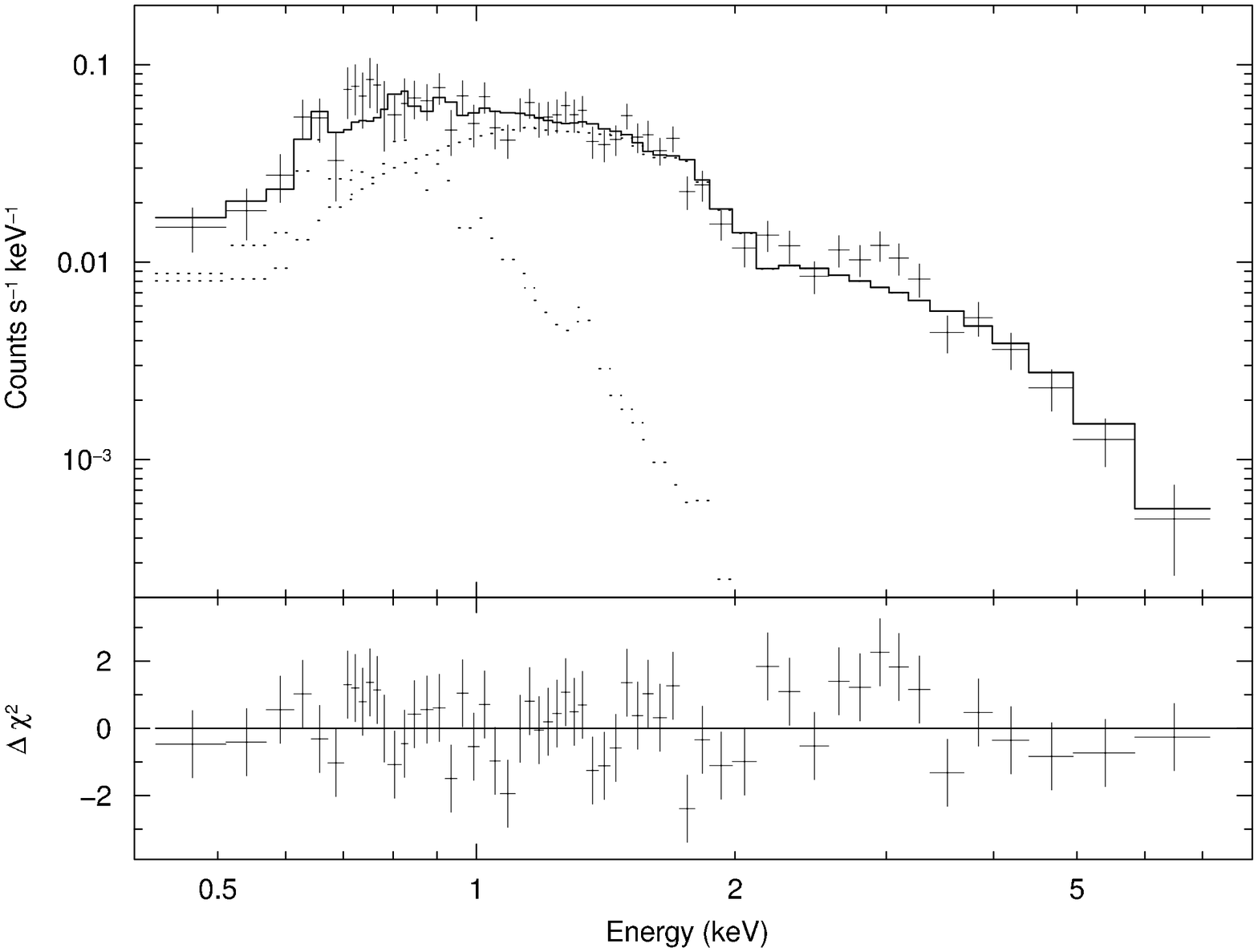}
\end{center}
\caption{Combined \CHANDRA\  spectra of point sources in the north ({\it left}) and south ({\it right})
regions. The solid lines show the best-fit absorbed 2T models,
with a Gaussian for the north spectrum.
The dotted lines show each plasma component. The bottom panels show residuals from
the best-fit models.
\label{fig:cxopointsource}
}
\end{figure*}

The \CHANDRA\  observation we analyzed was performed with the ACIS-I detector
on 1999 Sep. 6 for 21.5~ksec (sequence ID: 50, 1249),
and has about twice the exposure of the data set used in \citet{Evans2003}.
In this observation,
the ACIS-I covered most of the XIS \FOV\ (see Figure~\ref{fig:xmmsuzakuimage}).
We used the CIAO ver. 3.3 for the analysis.
Using {\tt wavdetect} on the 0.5$-$8~keV image and setting the detection
threshold to a significance 10$^{-6}$,
we detected 49 and 25 sources in the north and south regions respectively.
We extracted X-ray events of these point sources and
background from a nearby source free region.
Figure~\ref{fig:cxopointsource} depicts the composite spectra of the point sources
in the north and south regions.
We generated response matrices and auxiliary files with {\tt mkrmf} and {\tt mkwarf}.
The north spectrum shows weak neon and magnesium features.  In contrast, 
the south spectrum appears featureless, 
but this could be due to low spectral resolution of this data taken in an early observing phase
without CTI correction.
Each spectrum was fitted using an absorbed 2-temperature (2T) model
(WABS, \cite{Morrison1983}; the APEC\footnote{http://cxc.harvard.edu/atomdb/sources\_apec.html} code,
Table~\ref{tbl:cxopointsrc}).
A model of the south spectrum is reproduced at above 90\% confidence, while
a model of the north spectrum is reproduced at slightly less than 90\% confidence,
possibly because of marginal line features around 1~keV.
Both best-fit models have high \KT\  $\sim$4~keV and consistent \NH\  values with the interstellar absorption to the Carina Nebula.
The extremely low abundances are artifacts of an incompletely calibrated response
for the early \CHANDRA\ data and/or combining spectra from multiple sources.

The ACIS-I \FOV\  does not cover $\sim$15\% of the north region and $\sim$37\%
of the south region.
We therefore scaled the normalization of the spectra shown in Figure~\ref{fig:suzakuspecwcontami}
to account for the partial coverage, though in the \XMM\  image 
only a few point sources are detected in the uncovered portion of the north missing region and none in the south.
The estimate still has an uncertainty in that variability of point sources must be taken into account, though we do not expect sources to vary in concert.
The point source contribution to the \SUZAKU\  to the north and south regions 
inferred from our analysis of the \CHANDRA\ data
would be $\lesssim$1/3 of the total flux.
\citet{Townsley2006} suggests that the number of faint X-ray sources around the Tr~14 cluster could exceed one thousand; it is therefore possible that the contribution from faint point sources is
underestimated.

\begin{table*}
\begin{center}
\caption{Best-fit Models of Chandra Point Source Spectra \label{tbl:cxopointsrc}}
\begin{tabular}{lccc}
\hline\hline
Parameters&&North&South \\ 
\hline
\KT$_{cool}$&(keV)&0.62&0.27\\
log \EM$_{cool}$\footnotemark[$*$]&(\UNITEI)&55.7&55.9\\
\KT$_{hot}$&(keV)&3.8&4.2\\
log \EM$_{hot}$\footnotemark[$*$]&(\UNITEI)&55.6&55.6\\
Abundance  &(solar)&0.16&0.08\\
\NH&(10$^{21}$~\UNITNH)&1.3&1.7\\
Flux\footnotemark[$\dagger$]&(10$^{-13}$~\UNITFLUX)&9.0&7.5\\ \hline
$\Delta \chi^{2}$/d.o.f (d.o.f)&&1.40~(63)&1.20~(51)\\ \hline
\multicolumn{4}{@{}l@{}}{\hbox to 0pt{\parbox{100mm}{\footnotesize
\footnotemark[$*$] Assuming $d$ =2.3~kpc, the distance to $\eta$~Carinae \citep{Hillier2001}.
\par\noindent
\footnotemark[$\dagger$] Observed X-ray flux between 0.5--10~keV.
}\hss}}
\end{tabular}
\end{center}
\end{table*}

\subsection{CXB, GRXE, LHB and \etacar}
\label{subsec:cxbgrxehlbetacar}

For a CXB spectrum we use model Id1 in Table 2 of \citet{Miyaji1998},
which suggests a CXB flux fluctuation between regions of $\lesssim$30\%.
We assumed \NH\  at 1.3$\times$10$^{22}$~\UNITNH, the Galactic H\emissiontype{I}
column density to the Carina Nebula \citep{Dickey1990}.
The CXB flux contribution to the north and south spectra is $\lesssim$10\%.

To estimate the GRXE flux around the Carina Nebula ($l$, $b$) $\sim$(287.\DEGREE60, $-$0.\DEGREE63)
we referred to a 3--20~keV Galactic plane map
taken with Rossi X-ray Timing Explorer (RXTE) 
(see the top panel of Figure~7 in Revnivtsev et al. 2006).
In this map, the Carina Nebula is heavily contaminated by emission from \etacar\  
because of the limited RXTE spatial resolution.
We therefore measured X-ray flux at slightly higher ($l$ $\sim$290\DEGREE) and lower ($l$ $\sim$285\DEGREE)
Galactic longitudes along the Galactic plane, where bright X-ray point sources are 
apparently absent, and
interpolated those values to infer a GRXE flux at the Carina Nebula 
of $\sim$1.4$\times$10$^{-11}$~\UNITFLUX~deg$^{-2}$ (3$-$20~keV).
We assumed the GRXE spectral shape measured by \citet{Ebisawa2005}
at ($l$, $b$) $\sim$(28.\DEGREE5, 0.\DEGREE0) 
and reduced the normalization by a factor of 5 to match the GRXE flux at the Carina Nebula.
The GRXE emission contributes approximately the same flux as the CXB between 1 and 8~keV, $\lesssim$10\% of the total.

Emission from LHB toward the Carina Nebula has an insignificant spatial variation
over degrees, and has a surface brightness of $\sim$4$\times$10$^{4}$ counts s$^{-1}$ arcmin$^{-2}$ 
in the ROSAT PSPC R1+R2 band \citep{Snowden1998}.
Assuming a Raymond-Smith thin-thermal plasma model \citep{Raymond1977},
the LHB emission with \KT\  $\sim$0.1~keV accounts for 
$\sim$30\% of the 0.2$-$0.3~keV flux.
Since emission from the Carina Nebula should be cut off around $\sim$0.4~keV by interstellar absorption,
the residual emission below 0.4~keV probably also arises in the foreground.

Emission from the star \etacar\  is an order of magnitude stronger than any other source in the field
and thus can contaminate a substantial fraction of the XIS \FOV.
We reproduced the \etacar\ spectrum using the same \SUZAKU\  data, modeling it with
a commonly absorbed 2T model with an independent nitrogen abundance
for the soft diffuse emission below $\sim$1~keV
and a commonly absorbed 2T model for the central point source.
We then estimated effective area of a source at the position of \etacar\  on each source region
using the arf generator {\tt xissimarfgen-2006-05-28} (Figure~\ref{fig:suzakuspecwcontami}).
The contamination is almost negligible for the south region and $\lesssim$10\% for the north region.

\subsection{Total Contribution}

Figure~\ref{fig:suzakuspecwcontami} depicts the total contribution of potential contaminants
to the \SUZAKU\  spectra.
Below 0.3~keV, the LHB accounts for  $\sim$30\% of the observed flux.
Above $\sim$2~keV, the contamination sources contribute up to half of the
\SUZAKU\  flux.
This, in turn, means that half of the hard band flux cannot be attributed to the known contaminants.
On the other hand,
the excess between 0.3 and 2~keV is a factor of 3$-$10 higher than the total of possible contaminants
and therefore seems to originate in diffuse plasma
though a detailed analysis of the point source population would be desirable.

\begin{figure*}
\begin{center}
\FigureFile(80mm,53.7mm){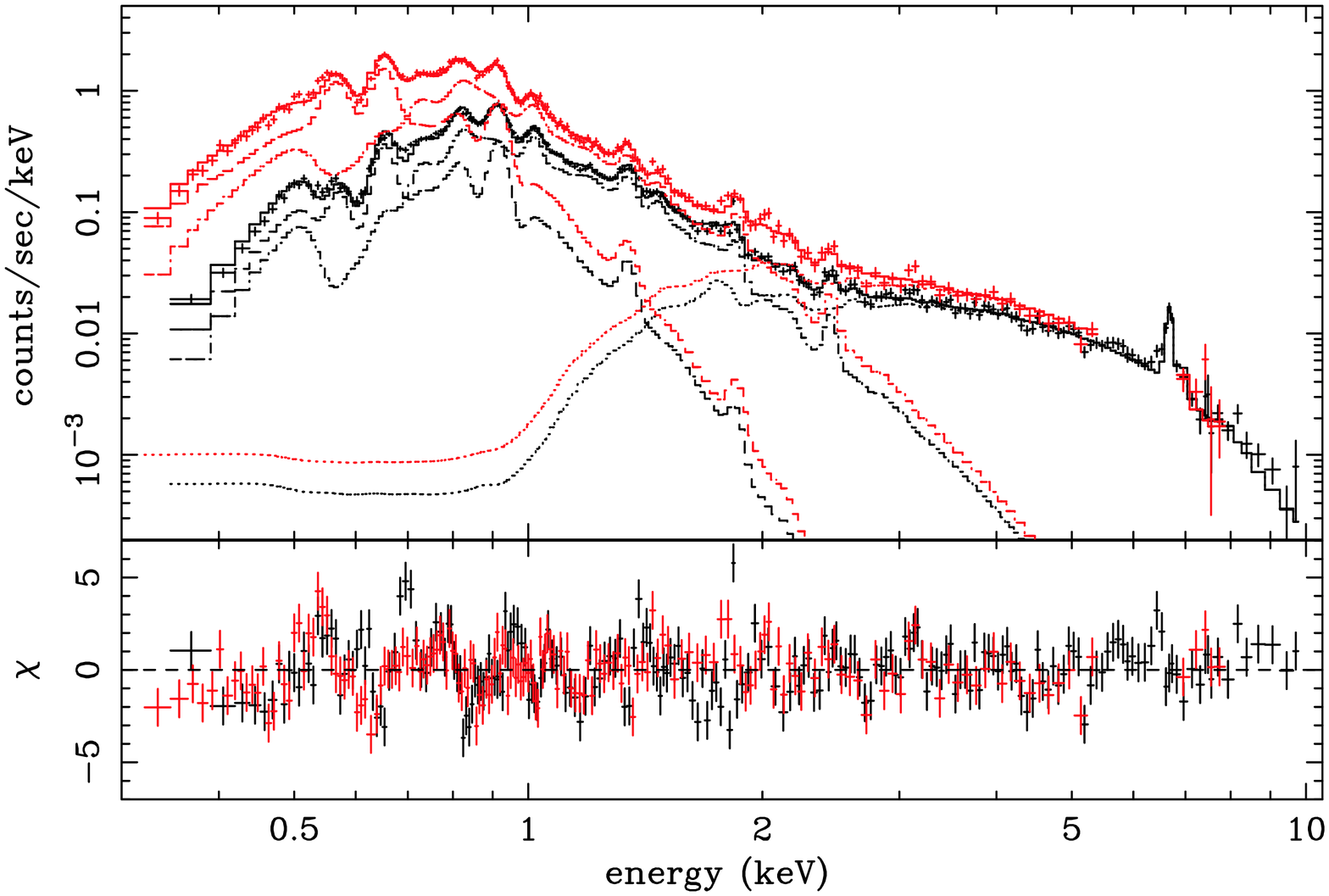}
\FigureFile(80mm,53.7mm){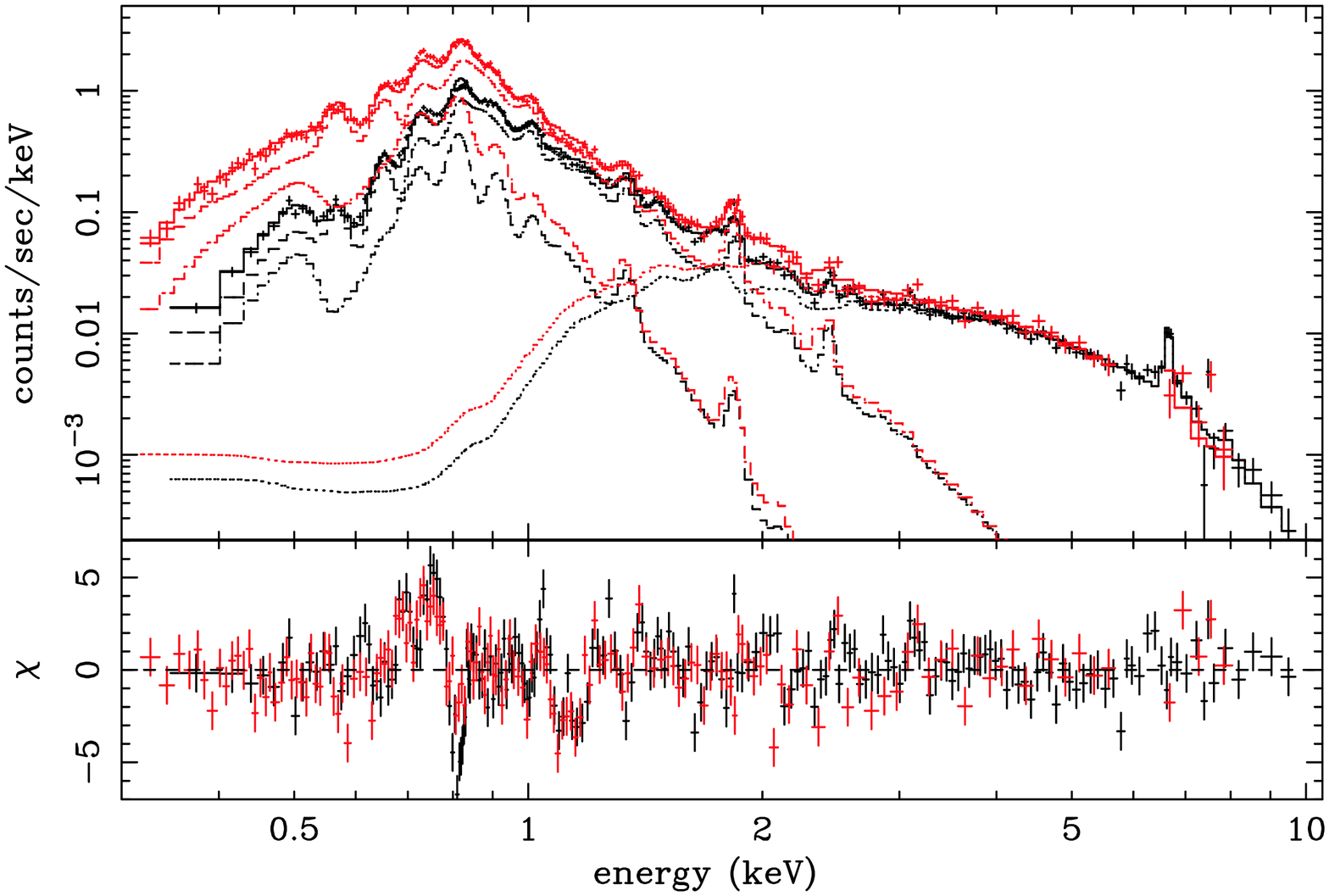}
\end{center}
\caption{Best-fit result of the north ({\it left}) and south ({\it right}) spectra
by the model with \NH(MC) and \KT\  tied ({\it black}: FI, {\it red}: BI).
The solid lines and dotted lines show the best-fit model and each of the plasma components, respectively.
The bottom panels show residuals in the fits.
\label{fig:suzakuspecfits}
}
\end{figure*}

\section{Spectral Fits}
\label{sec:diffuse}

To quantitatively estimate the spectral parameters of the diffuse emission,
we simultaneously fitted the FI and BI spectra above 0.3~keV for each region.
We tentatively assumed an absorbed 1T thin-thermal plasma 
(WABS, \cite{Morrison1983}; APEC code) for the HC,
which should really originate from multiple emission components as seen in \S\ref{sec:contami}.
For the MC,
we first assumed an absorbed 1T thin-thermal plasma, which however failed to
reproduce both the spectral slope above 1~keV and the strong emission line from O\emissiontype{VII} ions.
We therefore assumed a commonly absorbed 2T thin-thermal plasma (MC$_{cool}$ and MC$_{hot}$).
We omitted data below 0.3~keV, that is the SC, and 
BI data points between 5.5$-$6.7~keV, where emission from $^{55}$Fe calibration source dominates.
We separately varied elemental abundances between the HC and MC.
We also individually varied the abundance of each element of the MC,
but tied the abundances of the same elements between MC$_{cool}$ and MC$_{hot}$.

Calibration uncertainty is still a concern at the time of writing this paper.
To adjust the effective area between the XIS sensors
we varied the BI normalization by multiplying an energy independent coefficient
to the BI model (``constant" model in {\tt xspec}).  To allow for uncertainty in the absolute energy scale of each spectrum we allowed the gain to vary (``gain fit" function in {\tt xspec}).
In the best-fit models, the BI spectra of the north and south regions have $\sim$11\% and $\sim$5\% 
higher normalization than the FI spectra, respectively.
Other \SUZAKU\  observations also show $\sim$3\% normalization inconsistency between 
the BI and FI (e.g., NGC 2110, Bamba et al., private communication).
A remaining $\sim$5\% discrepancy in the north spectra may be caused by
inclusion of the $^{55}$Fe calibration source region for XIS1
($\sim$4\% discrepancy in the extracting area).
The slope and offset of the energy scale settled 
around $\sim$2.1\% and 12~eV for the BI south spectrum and
$\sim$0.1$-$0.4\% and $\sim$1$-$7 eV for the others,
values generally consistent with other rev 0.7 data.
In subsequent fits,
we fixed these parameters at their best-fit values.

The best-fit models (Table~\ref{tbl:specallfit}) reproduce neither of the spectra with a goodness of fit acceptable
above 90\% confidence.
This might be caused by poor calibration and/or inaccuracy in the emission line code,
which we describe in detail later.
In this model, plasma temperatures of both spectra are similar to each other, but 
the north spectrum has an order of magnitude larger cool component emission measure \EM(MC$_{cool}$)
and a factor of two larger \NH(MC) than the south spectrum.
Fits of \CHANDRA\ spectra also showed larger \NH\ to the north of the diffuse plasma 
($\sim$7$\times$10$^{21}$~\UNITNH) than a negligible \NH\ to the south
\citep{Townsley2006}.
However,
factor of 2 larger \NH\  value for the north region seems inconsistent 
with a thick CO condensation being around the south region \citep{Yonekura2005},
though we do not know locations of the X-ray plasma and the CO condensation along the line of sight.
Moreover, it would be too convenient to think that the north region happens to have an order of magnitude larger \EM(MC$_{cool}$), and, as a result, have a 0.3$-$0.5~keV observed flux 
similar to the south region.
Since \EM(MC$_{cool}$) and \NH(MC) can easily couple in a spectral fit,
it is possible that the differences of \EM(MC$_{cool}$) and \NH(MC) arise artificially as the result of a local $\chi^{2}$ minimum.

We therefore assumed the north and south regions have the same absorption and 
tied together \NH(MC).
We also tied together the temperatures \KT(MC$_{cool}$), \KT(MC$_{hot}$) and \KT(HC), based on their similarity in the unconstrained fits.
The best-fit model (Figure~\ref{fig:suzakuspecfits}, Table~\ref{tbl:specallfit})
is once again not acceptable above 90\% confidence, 
and the residuals are similar to these in the individual fits.
A broad residual hump is visible around 0.65$-$0.80~keV,
where emission from the Fe-L shell complex is strong.
The hump is not reduced using a multi-temperature model (c6pvmkl) in {\tt xspec}.
This may mean that the plasma is not in collisional equilibrium, 
or that the emission line code does not reproduce the Fe L-shell line complex.
On the other hand,
the model does not reproduce the argon and sulfur lines around 2$-$3~keV, 
confirming that the HC has multiple components
(see \S\ref{sec:contami}).
Both spectra have marginal peaks with equivalent width of $\sim$50$-$60 eV at 6.4~keV,
which might arise from fluorescence of cold iron.

\begin{table*}
\begin{center}
\caption{Spectral Fitting Results\label{tbl:specallfit}}
\begin{tabular}{ccccccccccc}
\hline\hline
Model&&\multicolumn{2}{c}{Individual}&\multicolumn{2}{c}{\NH(MC), \KT\  tied}&\multicolumn{2}{c}{\NH(MC) fixed}&\multicolumn{2}{c}{NEI}&Typical Error\footnotemark[$*$]\\
&&North&South&North&South&North&South&North&South\\
\hline
\multicolumn{4}{l}{Medium Component (MC)}\\
\KT$_{cool}$&(keV)&0.20&0.23&\multicolumn{2}{c}{0.21}&\multicolumn{2}{c}{0.19}&\multicolumn{2}{c}{0.21}&0.01\\
\KT$_{hot}$&(keV)&0.60&0.58&\multicolumn{2}{c}{0.58}&\multicolumn{2}{c}{0.56}&\multicolumn{2}{c}{0.57}&0.01\\
log \EM$_{cool}$\footnotemark[$\dagger$]&(\UNITEI arcmin$^{-2}$)&56.0&55.0&55.7&55.6&56.2&56.0&55.5&55.3						&0.1\\
log \EM$_{hot}$\footnotemark[$\dagger$]&(\UNITEI arcmin$^{-2}$)&55.0&54.9&55.1&55.0&55.0&54.8&55.1&55.0						&0.1\\
\NH&(10$^{21}$~\UNITNH)&2.4&1.2&\multicolumn{2}{c}{1.8}&\multicolumn{2}{c}{3.0 (fix)}&\multicolumn{2}{c}{1.8}&0.2\\
C  &(solar)&2.5e-2&0.0&0.0&0.0&0.34&0.85&0.0&0.25													&5e-2\\
N  &(solar)&0.0&1.2e-2&0.0&0.0&0.0&0.0&1.5e-2&4.5e-2												&1e-2\\
O  &(solar)&4.3e-2&0.12&6.2e-2&6.3e-2&4.2e-2&6.1e-2&9.1e-2&0.12										&1e-2\\
Ne &(solar)&8.9e-2&0.14&0.13&9.0e-2&8.1e-2&0.11&0.19&0.14  											&3e-2\\
Mg &(solar)&0.11&0.19&0.11&0.16&9.4e-2&0.17&0.11&0.16												&2e-2\\
Al  & (solar)&0.17&0.0&7.5e-2&2.9e-2&0.17&0.18&&    										&0.2\\
Si &(solar)&0.10&0.25&9.2e-2&0.24&0.11&0.36&0.11&0.27												&2e-2\\
S & (solar)&0.38&0.12&0.24&0.39&0.48&0.85&0.22&0.43												&0.1\\
Ar&(solar)&0.36&0.0&0.0&0.13&1.4e-2&1.2&1.4e-2	&1.2													&0.2\\
Ca&(solar)&0.0&0.0&0.0&0.0&0.0&0.0&0.0&0.0														&3e-2\\
Fe &(solar)&8.4e-2&0.30&7.8e-2&0.30&8.3e-2&0.43&8.3e-2&0.30											&1e-2\\
Ni &(solar)&0.34&0.56&8.9e-2&0.78&0.56&1.9&1.5e-2&0.39												&0.1\\
$\tau$&(s cm$^{-3}$)&&&&&&&\multicolumn{2}{c}{1.7e12}												&$>$ 9e11\\
\multicolumn{4}{l}{Hard Component (HC)}\\
\KT&(keV)&5.5&5.5&\multicolumn{2}{c}{5.3}&\multicolumn{2}{c}{5.5}&\multicolumn{2}{c}{5.1}						&0.2\\
log \EM\footnotemark[$\dagger$]&(\UNITEI arcmin$^{-2}$)&54.4&54.5&54.4&54.5&54.4&54.5&54.5&54.5								&0.1\\
\NH&(10$^{22}$~\UNITNH)&2.3&1.3&2.4&1.5&1.8&1.2&2.5&1.7  											&1\\
Z&(solar)&0.48&0.44&0.47&0.43&0.50&0.44&0.46&0.42													&5e-2\\ \hline
$\chi^{2}/d.o.f$&&2.07&2.94&\multicolumn{2}{c}{2.70}&\multicolumn{2}{c}{3.52}&\multicolumn{2}{c}{2.70}\\
$d.o.f.$&&378&344&\multicolumn{2}{c}{736}&\multicolumn{2}{c}{737}&\multicolumn{2}{c}{737}\\ \hline
\multicolumn{11}{@{}l@{}}{\hbox to 0pt{\parbox{180mm}{\footnotesize
\footnotemark[$*$]The typical errors show the range within $\Delta\chi^{2} =$2.7.
\par\noindent
\footnotemark[$\dagger$]Assuming $d$ =2.3~kpc.
}\hss}}
\end{tabular}
\end{center}
\end{table*}

Although the best-fit models are not formally acceptable,
they seem to reproduce the global spectral shapes well.
The plasma temperatures, $\sim$0.2 and $\sim$0.6~keV in the MC
and $\sim$5.5~keV in the HC, are similar to the results of the individual fits.
The \EM(MC$_{cool}$) and \EM(MC$_{hot}$) have the same ratio in the north and south spectra,
confirming that the two spectra are quite similar between 0.3 and 2~keV.
The average surface fluxes of the north and south spectra are 
7 and 9.5 $\times$10$^{-14}$~\UNITFLUX\  arcmin$^{-2}$ between 0.5$-$2~keV
and 3.5 and 3.8 $\times$10$^{-14}$~\UNITFLUX\  arcmin$^{-2}$ between 2$-$8~keV, respectively.
The 0.5$-$2~keV band surface flux is consistent with the unabsorbed 
surface flux estimated from the least 
contaminated \CHANDRA\  spectrum, assuming \NH\  =3$\times$10$^{21}$~\UNITNH\  \citep{Evans2003}.
This result ensures that the \SUZAKU\  spectra between 0.3$-$2~keV represent diffuse emission
from the Carina Nebula.

The south region has a factor of 2 higher Si abundance and a factor of 4 higher 
iron abundance than the north region.
Abundances of magnesium and nickel may also be enhanced by a factor of $\sim$1.5 and 7$-$8 
in the south,
though these lines are situated among the iron L-shell lines
and their abundance measurements are strongly affected by the remaining uncertainty
in fitting the iron line complex.
No other elements show a significant abundance difference.

The derived \NH\  ($\sim$1.8$\times$10$^{21}$~\UNITNH) is consistent with \NH\ 
to some objects around the Carina Nebula \citep{Seward1979,Savage1977}.
However, Leutenegger, Kahn and Ramsay (2003) argued that these measurements are problematic and suggested
a factor of 1.5 higher \NH\  ($\sim$3$\times$10$^{21}$~\UNITNH).
When we fixed \NH\  at 3$\times$10$^{21}$~\UNITNH\  in the spectral fit,
the abundances of silicon, sulfur and carbon,
which have emission lines in the 0.4$-$0.5~keV band where the spectrum cuts off 
sharply due to interstellar absorption, significantly increase
(see Table~\ref{tbl:specallfit}).
This means that abundances of these elements strongly couple with \NH, 
and are therefore less reliable.

Since some emission lines such as Mg\emissiontype{XI} and Fe\emissiontype{XVII} have
residuals on the low energy side,
we also tried a non-equilibrium ionization (NEI) model (Table~\ref{tbl:specallfit}).
However, the large ionization time scale $\tau \sim$1.7$\times$10$^{12}$~s~cm$^{-3}$ in the best-fit 
rather suggests the ionization equilibrium plasma, and therefore
the best-fit model did not reduce the residuals.
We also tried multi-temperature models for the cool component, but the result did 
not improve, either.

\section{Discussion}
\label{sec:discussion}

Based on the morphology of the X-ray emission, the X-ray luminosity ($\sim$10$^{35}$~\UNITLUMI)
and the plasma temperature ($\sim$0.8~keV),
\citet{Seward1982} argued that diffuse X-ray emission from the Carina Nebula is probably produced 
by strong stellar winds from early stars 
colliding with ambient gas.
Our results using \SUZAKU\  and \XMM\  are basically consistent with these previous results,
except for a slightly lower temperature ($\sim$0.2, 0.6~keV).
The inferred temperature is also similar to the temperature of the diffuse emission
of the Omega and Rosette Nebulae, for which \citet{Townsley2003} suggested a stellar wind origin.

What is new in this result is the detailed line diagnostics
in the soft energy band, especially below $\sim$1~keV.
The study reveals that nitrogen is significantly depleted relative to other heavy elements throughout the observed regions,
while the abundance of iron and silicon in the southern portion of the nebula is 2-4 times 
higher than in the northern portion.
Any successful explanation of the diffuse emission from the Carina Nebula must account for these abundance anomalies.  
In the paragraphs below, we explore possible mechanisms.

The evolved stars in the field, \etacar\  and WR~25, have strong winds rich in nitrogen and 
poor in oxygen,
resulting from the CNO hydrogen burning process \citep{Davidson1982,Hucht1981}.
Strong nitrogen and negligible oxygen lines are observed from the diffuse X-ray 
plasma immediately surrounding \etacar\ ;
the abundance ratio was restricted to N/O $>$9 
(\cite{Tsuboi1998,Leutenegger2003}; 
and the \etacar\ spectrum from the same \SUZAKU\ observation).
Although less evolved massive O stars in general show a nitrogen overabundance, the O stars in the 
Carina region have a relatively low N/O ratio \citep{Morrell2005}.
Even so, the \SUZAKU\  spectra of the Carina Nebula show no significant nitrogen emission,
but show clear oxygen emission lines.
The N/O abundance ratio inferred from the spectral fits is $\lesssim$0.4,
over 20 times less than around \etacar.
The abundance distribution is totally contrary to that expected from massive stellar winds,
unless the winds somehow heat the interstellar matter without enriching it, thus leaving the 
X-ray plasma with abundances typical of interstellar matter.
At the same time,
X-ray luminosity of the Carina Nebula is about two orders of magnitude higher than that
of other Galactic star forming regions, but the number of early O stars is only an order of magnitude higher
(see Table~4 in \cite{Townsley2003}).
These results suggest an additional energy source is needed to power the X-ray emission in the Carina Nebula.

An obvious possibility is one or more core-collapse supernovae (i.e. Type Ib,c or II), 
mentioned as a possibility by \citet{Townsley2003}.
The north and south regions vary strongly in silicon and iron abundances and
marginally in magnesium and nickel abundances.
These elements are products of core-collapse supernovae.
Moreover, young SNRs such as Cas A and Vela show strong abundance variation from location 
to location (e.g., \cite{Katsuda2006}).

A single supernova could power the entire Carina Nebula.  Assuming a uniform temperature throughout the entire extent of the nebula (1.\DEGREE1 E-W by 0.\DEGREE7 N-S \citep{Seward1982} by a comparable line-of-sight depth) the total energy content in the hot gas of $\sim$2$\times$10$^{50}$ ergs, a modest fraction of the $\sim$10$^{51}$ ergs of kinetic energy produced by a canonical supernova.  

On the other hand, the total iron mass in the diffuse gas cannot be the product of a single supernova.
Assuming an iron abundance of 0.30~solar
throughout the entire nebula implies a total iron mass $M_{Fe} \sim$ 0.1/$n$ \UNITSOLARMASS, where $n$ is the plasma number density in the unit of cm$^{-3}$.
Our results suggest  $n \sim $0.2$-$0.4 cm$^{-3}$ in the north and south regions,
assuming the diffuse plasma in each region has a dimension along the line of sight comparable to its project angular extent at a distance of 2.3~kpc.  If these regions are representative of the entire nebula then $M_{Fe} \sim$0.25$-$0.4~\UNITSOLARMASS.  The typical core collapse supernova yields $\le$0.1~\UNITSOLARMASS, so more at least 3-5 supernovae would be required to supply the iron.  

There is no evidence for supernova remnants in the Carina Nebula in the radio \citep{Whiteoak1994} or the X-ray.  There are two known pulsars, 1E 1048.1-5937 and PSR J1052$-$5954, 
within 1\DEGREE\  from \etacar,
but they are located far behind the Carina Nebula \citep{Kramer2003,Gavriil2004,Gaensler2005}.  Core collapse supernovae produce a wide variety of stellar remnants, however, and not all of them pulse.  It is plausible that several stellar remnants similar to those in Cas A and Puppis A (with \LX\  $\sim$2$\times$10$^{33}$~\UNITLUMI\  and \KT\  $\sim$5$\times$10$^6$ K) could reside undetected in the Carina Nebula.  There is no requirement that the stellar remnants are located within the north or south regions for which we carefully inspected the point sources, or even in the Carina Nebula after $\sim$10$^6$ yr.  

The need for multiple supernovae to explain the abundances with the modest energy input requirements is consistent with a hypothesis advanced by \citet{Yonekura2005} to account for the fact that the 
strong turbulence in the 
\etacar\  giant molecular cloud (GMC) observed in CO lines is too energetic to be produced by
stellar winds or a single SNR.  They suggest that the turbulence was preexisting when the GMC was formed.  They postulate that it was produced by the large number of supernovae ($>$20) responsible for creating the Carina flare supershell \citep{Fukui1999}, and is now dissipating.  
On the other hand, \citet{Smith2000} found a bipolar morphology in the Carina Nebula 
in the infrared and suggested the presence of one or more supernova remnants.
Whichever case is connected to the diffuse X-ray emission,
the absence of evidence of SNRs in the Carina Nebula is not a problem, as the explosions that power the X-ray nebula (and the molecular cloud turbulence) occurred $\sim$10$^{6}$ years ago.  

The abundance variation between the north and south regions could be one vestige of the multiple explosions.  Alternatively, since silicon, iron and magnesium are refractory elements required for silicate grains and can be 
very strongly depleted from the gas phase  in the ISM \citep{Salpeter1977,Mathis1996,Draine2003},
we may consider the dust formation or destruction,
both of which are strongly associated with the environment of evolved massive stars,
to explain the abundance variation.

SNRs thus seem to be a promising hypothesis for explaining the diffuse plasma
from the Carina Nebula.
However, there are several issues yet to be addressed to confirm this hypothesis.
(i). The absolute abundances of all the heavy elements are smaller than solar,
unlike in a canonical supernova remnant.
Other \SUZAKU\  observations of diffuse plasma, including the Cygnus Loop supernova remnant
\citep{Miyata2006}, also show similar low absolute elemental abundances.
This could be simply due to a problem of the current X-ray measurement,
such as the need to include a non-thermal continuum component, unknown physical processes of line emission 
in a very low density plasma, or helium-rich plasma.
(ii). The oxygen abundance is relatively low though it should be the most abundant element 
in core collapse supernova ejecta.
In particular, the O/Fe ratio is smaller than the solar ratio.

What is the excess below the O\emissiontype{VII} line in the north spectrum?
Since the other emission lines did not show such an apparent broadening,
poor energy calibration is an unlikely cause.
The excess can be reproduced by a narrow Gaussian line with a center energy 
of 0.547$\pm$0.01~keV.
The error range includes three emission lines from S\emissiontype{XIV} 
(10$\rightarrow$1, 11$\rightarrow$1) and Ca\emissiontype{XVI} (95$\rightarrow$8).
However,
sulfur and calcium should have stronger peaks below $\lesssim$0.5~keV
when \KT\  $\sim$0.1$-$0.8~keV, which are not seen in the spectrum.
Though strong radiation from Tr~14 cluster and \etacar\  could photo-ionize the plasma,
the excess peak is slightly lower than the forbidden line of O\emissiontype{VII} at 0.561 keV,
which is strong in the photo-ionized plasma \citep{Porquet2000}.
If the excess is produced by the Doppler shift of the helium-like oxygen line,
the velocity is extremely high ($\sim$7.5$\times$10$^{3}$~\UNITVEL).
We have no good explanation for this feature.
We should note that 
a similar excess below O\emissiontype{VII} is seen from a few \SUZAKU\  observations
such as the North Polar Spur (Miller et al. in preparation).

Less than half of the surface flux from the hard component  arises from known contaminants: X-ray point
sources, the CXB and the GRXE.
One possible origin of the remainder is a large number of 
low mass young stars embedded in the cloud, which are too faint to be detected 
individually.
Both the north and south regions include massive clouds \citep{Yonekura2005},
which probably have the star forming activity.
Low mass pre-main-sequence stars in the Orion Nebula emit X-rays of \KT $\sim$0.5$-$3~keV
and \LX\ $\sim$10$^{28.9}$~\UNITLUMI\  between 2$-$8~keV \citep{Feigelson2005}.
Since 2$-$8~keV X-ray surface brightness of the Carina Nebula is 2.3$\times$10$^{31}$~\UNITLUMI~arcmin$^{-2}$,
the surface number density of young stars should be $\sim$290~arcmin$^{-2}$ = 650~pc$^{-2}$ at 2.3~kpc.
Since the star formation would occur over a parsec scale along the line of sight, the number density 
should be $<$650~pc$^{-3}$, which is small compared with the Orion Nebula
(peak of $>$10$^{4}$ pc$^{-3}$, \cite{Hillenbrand1997}).
Multiple X-ray sources detected in a deep \CHANDRA\ observation of the Tr~14 cluster 
\citep{Townsley2006} may account for a part (or all) of this possible hidden population.
There are two problems:
hard X-ray surface flux is almost the same in the two regions even though there are more potential star formation sites in the north; and
\KT\  $\sim$5.5~keV is higher than the typical temperature
of low mass young stars
though a possibly different population in the Carina Nebula might produce a composite
spectrum with a higher effective plasma temperature.
Another possibility is that the hard X-ray emission comes from diffuse plasma.
Such a very high temperature plasma perhaps needs young supernova remnants 
(e.g., \cite{Koyama2006b}),
though the observed soft X-ray emission rather suggests presence of old supernova remnants.

\section{Summary}

We analyzed diffuse X-ray emission from the Carina Nebula using the
CCD camera XIS onboard the \SUZAKU\  observatory.
The XIS has the best soft band spectral resolution for extended sources of any X-ray observatory,
and provides detailed line profiles and reliable elemental abundances of the plasma.
To understand spatial distribution of the diffuse emission,
we first analyzed an \XMM\  MOS image centered at \etacar.
An X-ray color map indicates that the
diffuse emission north of \etacar\  is softer than the emission to the south.
Analysis of the \SUZAKU\  spectra shows that this difference arises from 
a difference in the iron L-shell line intensity.
\SUZAKU\  spectra are roughly divided into three X-ray components.
Spectra below $\sim$0.3~keV have almost the same surface flux in the north and south
regions.
This component originates in the foreground, including the LHB.
The spectra above 2 keV also have almost the same surface flux and shape. 
About half of the emission arises from the CXB, GRXE and X-ray point sources
detected with \CHANDRA.
The remainder of the hard emission may be from unresolved pre-main-sequence stars.
The spectra between 0.3 and 2~keV are dominated by emission from soft diffuse emission.
The spectra show K-shell lines of silicon, magnesium, neon,
oxygen, and possibly carbon ions, and L-shell lines of iron ions.
The strength of the Si and Fe lines is different between the north and
south regions, while the other parts of the spectra are similar.
The spectra from both regions are reproduced with a 2T plasma model with
\KT\  $\sim$0.2 and 0.6~keV and sub-solar elemental abundances
and common absorption of $\sim$2$\times$10$^{21}$~\UNITNH,
but are not acceptable at the 90\% confidence level.
No nitrogen line is detected from either of the spectra, restricting N/O $\lesssim$0.4.
Such a low nitrogen abundance is not expected from stellar winds from evolved massive stars,
and suggests that the plasma originates from an old SNR, or a super shell
produced by multiple SNRs.
The south region has a factor of 2$-$4 higher elemental abundances in
iron and silicon than the north region.
Magnesium and nickel may show spatial abundance variation, as well.
The abundance variation may be explained by the presence of SNR ejecta.
Alternatively, dust formed around the star forming core may play an
important role.

The spatial structure and abundance variation of the plasma provide clues to 
the origin of the diffuse emission.
For this purpose, a mapping of the entire Carina Nebula with \SUZAKU\  and \XMM\  will be important.
Meanwhile, the population of faint stars in the Carina Nebula should be investigated with a deep \CHANDRA\ observation, through a study of the X-ray luminosity function.
The spectra are complicated with many emission
lines; high resolution spectroscopy on the future satellites, such as an X-ray calorimeter, will be 
needed to solve the origin of this mysterious X-ray nebula.

\bigskip
The authors appreciate the comprehensive comments and suggestions by the referee, Dr. Leisa K. Townsley.
K.\, H. is financially supported by a US \CHANDRA\  grant No. GO3-4008A.
M.\,T. and Y.\, E. are financially supported by the Japan Society for the Promotion of Science.

\bibliographystyle{apj}
\bibliography{inst,sci_AI,sci_JZ,scibook}

\clearpage

\clearpage

\end{document}